\documentclass[a4paper]{article}

\usepackage[english]{babel}
\usepackage[utf8x]{inputenc}
\usepackage[T1]{fontenc}

\usepackage[a4paper,top=3cm,bottom=2cm,left=3cm,right=3cm,marginparwidth=1.75cm]{geometry}

\usepackage{setspace}
\usepackage{amssymb}

\usepackage{cite}

\usepackage{amsmath}
\usepackage{graphicx}
\usepackage[colorinlistoftodos]{todonotes}
\usepackage[colorlinks=true, allcolors=blue]{hyperref}

\title{Broadband switchable terahertz half-/quarter-wave plate based on VO$_2$-metal hybrid metasurface with over/underdamped transition}

\author{Xiaoqing Luo$^{1,2}$, Fangrong Hu$^{3}$, and Guangyuan Li$^{1,2,4,*}$}

\date{}
\begin{document}
\maketitle

\begin{spacing}{2.0}

\noindent \large$^1$CAS Key Laboratory of Human-Machine Intelligence-Synergy Systems, Shenzhen Institutes of Advanced Technology, Chinese Academy of Sciences, Shenzhen, 518055 China

\noindent  $^2$Guangdong-Hong Kong-Macao Joint Laboratory of Human-Machine Intelligence-Synergy Systems, Chinese Academy of Sciences, Shenzhen Institutes of Advanced Technology, Shenzhen 518055, China

\noindent  $^3$Guangxi Key Laboratory of Optoelectronic Information Processing, Guilin University of Electronic
Technology, Guilin 541004, China

\noindent  $^4$Shenzhen College of Advanced Technology, University of Chinese Academy of Sciences, Shenzhen 518055, China


\noindent *Corresponding author: gy.li@siat.ac.cn

\end{spacing}

\begin{abstract}
Dynamically switchable half-/quarter-wave plates have recently been the focus in the terahertz regime. Conventional design philosophy leads to multilayer metamaterials or narrowband metasurfaces. Here we propose a novel design philosophy and a VO$_2$-metal hybrid metasurface for achieving broadband dynamically switchable half-/quarter-wave plate (HWP/QWP) based on the transition from the overdamped to the underdamped resonance. Results show that, by varying the VO$_2$ conductivity by three orders of magnitude, the proposed metasurface's function can be switched between an HWP with polarization conversion ratio larger than 96\% and a QWP with ellipticity close to $-1$ over the broad working band of 0.8--1.2~THz. We expect that the proposed design philosophy will advance the engineering of metasurfaces for dynamically switchable functionalities beyond the terahertz regime.
\end{abstract}

\newpage

Polarization is an important characteristic of electromagnetic waves, and the manipulation of the polarization is desirable in a diverse range of applications, such as sensing, imaging, encryption, and communications \cite{FPC2010PolMM_rev}. Conventional approaches based on birefringence or total internal reflection effects are usually bulky and narrowband, and sometimes lossy. To circumvent these problems, metamaterials and their two-dimensional counterparts, metasurfaces, have recently been extensively explored because of their appealing merits such as compactness, high efficiency, broad bandwidth and easy integration \cite{FPC2010PolMM_rev,RPP2016Metasurface_Rev,AM2019ManipulMS_rev,NanoP2020PolMS_rev}.

In the terahertz regime, metamaterial- or metasurface-based polarization converters, especially HWPs and QWPs, are of particular interest due to the lack of suitable natural materials. Over the years, these terahertz metadevices have evolved from narrowband \cite{OE2009THzPolConvertNarrow} to broadband \cite{Science2013THzPolConvertBroad}, and from untunable to actively reconfigurable \cite{JPD2020TuneMM_rev,Res2020TuneM_rev,MOTL2020TuneMM_rev}. For example, based on metamaterials or metasurfaces that are made of metals or dielectrics, broadband terahertz HWPs \cite{APL2014HWP-THZ,EPL2017HWPQWPBroad,OME2017HWPBroad} or QWPs \cite{LPR2014QWP-THZ,SCIREP2016qwp-thz,OL2016QWPTHZ,EPL2017HWPQWPBroad,OE2018QWPHWP-THZ} have been proposed or demonstrated. Quite recently, dynamically switchable HWPs/QWPs based on metamaterials or metasurfaces incorporating tunable materials such as graphene \cite{OSAC2018QWP2HWP,OE2018SwitchQWP,PTL2019TunableQWP,OL2019HWP2QWP_GSi} or VO$_2$ \cite{SCIREP2015QWPVO2,IEEEPJ2016QWPVO2,OE2020HWP2QWP,CPL2020QWP2HWP} have received increasing attention. 

Although the structures for terahertz switchable HWP/QWP, as well as for other switchable bi-functionalities, such as switchable HWP and perfect absorber \cite{AOM2018HWP2ABS}, or switchable on/off asymmetric transmission \cite{OE2020switchableAT}, may be distinct in the literature, the design philosophy is almost the same. One first treats graphene with small Fermi level (or VO$_2$ in the insulating state at low temperature) as transparent dielectric because of the low conductivity, and design metal structures for one functionality such as HWP without considering graphene (or VO$_2$). For graphene with high Fermi level (or VO$_2$ in the conducting state at high temperature), the conductivity is as high as metal-like material, such that one can approximately treat graphene (or VO$_2$) as metal and then add graphene (or VO$_2$) structures to the metal structures for the other functionality such as QWP. This design philosophy usually results in metamaterials composed of stacked multilayer structures, complicating the fabrication and thus hindering the practical applications. A typical example is our previous design for achieving broadband switchable HWP/QWP \cite{OE2020HWP2QWP}, where the metal structures function as a broadband HWP while the VO$_2$ structures act as a broadband QWP. 

Following the above design philosophy, there exist a few exceptions \cite{SCIREP2015QWPVO2,IEEEPJ2016QWPVO2,OL2019HWP2QWP_GSi}, which are planar metasurfaces, greatly facilitating the fabrication compared with the multilayer metamaterials. However, their bandwidths are extremely narrow. For example, the switchable QWP based on a VO$_2$-metal hybrid metasurface works at only one \cite{SCIREP2015QWPVO2} or two specific frequencies \cite{IEEEPJ2016QWPVO2}, and the switchable HWP/QWP based on a graphene-dielectric hybrid metasurface has narrow bandwidth of 0.1~THz around 1~THz \cite{OL2019HWP2QWP_GSi}. In other words, it remains challenging to design a broadband switchable HWP/QWP based on the metasurface.

In this letter, we propose a VO$_2$-metal hybrid metasurface for achieving broadband switchable HWP/QWP in the terahertz regime. This metasurface is designed following a novel philosophy based on the transition from the overdamped to the underdamped resonance. We will show that by varying the VO$_2$ conductivity by three orders of magnitude, which can be done through thermal, optical or electrical stimulus \cite{NPGAM2018_VO2Rev}, the proposed metasurface functions as a switchable HWP/QWP with high efficiencies over a broad band of 0.8--1.2~THz.

Based on Jones matrix and Poincar\'e sphere analyses, Ma {\sl et al.} \cite{EPL2017HWPQWPBroad} developed a set of general criteria to design high-efficiency broadband terahertz waveplates using reflective metasurfaces:
\begin{equation}
    \label{eq:HWP}
    |r_{xx}|\approx |r_{yy}|\approx 1, \quad \& \quad {\rm arg} (r_{yy}) - {\rm arg} (r_{xx}) = 180^{\circ}\,,
\end{equation}
for an HWP, and for a QWP, 
\begin{equation}
    \label{eq:QWP}
    |r'_{xx}|\approx |r'_{yy}|\approx 1, \quad \& \quad {\rm arg} (r'_{yy}) - {\rm arg} (r'_{xx}) = \pm 90^{\circ}\,.
\end{equation}
Here $r_{xx}$ and $r_{yy}$ (or $r'_{xx}$ and $r'_{yy}$) are the reflection coefficients for the $\vec{E} \parallel \hat{x}$ and $\vec{E} \parallel \hat{y}$ polarizations, respectively, and the sign “$+/-$” denotes right- or left-circular final polarization state. 

Therefore, in order to realize a metasurface that can be switched from an HWP to a QWP, here we propose a novel design philosophy. We first design a metal/insulator/metal (MIM) configuration with $r_{xx}$ working in the underdamped regime whereas $r_{yy}$ in the overdamped regime, such that all the reflection coefficients have near-unitary amplitude, indicating low absorption, and that $r_{xx}$ can exhibit a 360$^\circ$ reflection phase coverage as frequency varies. This can lead to a high performance HWP according to Eq.~(\ref{eq:HWP}). As the conductivity of VO$_2$ (or graphene) $\sigma$ increases, $r_{xx}$ remains unchanged so that 
\begin{equation}
    \label{eq:rxx}
   |r'_{xx}|\approx |r_{xx}|, \quad \& \quad {\rm arg} (r'_{xx}) \approx {\rm arg} (r_{xx})\,,
\end{equation}
and meanwhile $r_{yy}$ transits from the overdamped to the underdamped regime so that 
\begin{equation}
    \label{eq:ryy}
    {\rm arg} (r'_{yy}) - {\rm arg} (r_{yy}) = -270^{\circ}\,.
\end{equation}
Satisfying Eqs.~(\ref{eq:rxx}) and (\ref{eq:ryy}) when $\sigma$ increases will result in dynamic switching from an HWP to a QWP. In other words, the metasurface initially has an underdamped and an overdamped resonances for two orthogonal polarizations, leading to an HWP, and as $\sigma$ increases the overdamped resonance transits into an underdamped resonance, resulting in a QWP.

\begin{figure}[htbp]
\centering
\includegraphics[width=8.8cm]{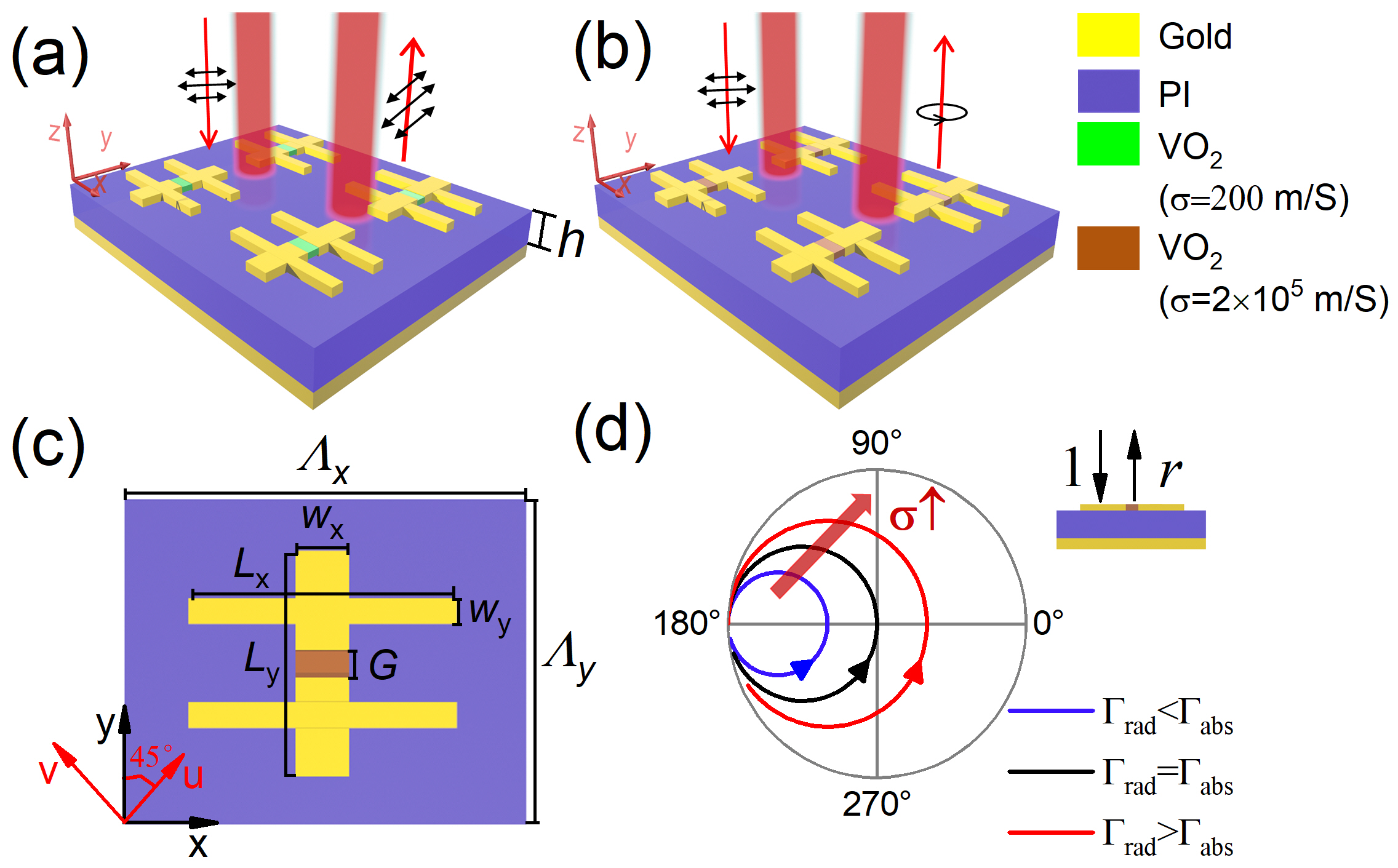}
\caption{(a)--(c) Schematics of the proposed VO$_2$-metal hybrid metasurface for achieving broadband switchable terahertz HWP/QWP. Geometric parameters used in this work are $\Lambda_{x} =162~\mu$m, $\Lambda_{y} =120~\mu$m, $h=33~\mu$m, $L_{x}=95~\mu$m, $L_{y}=80~\mu$m, $w_{x}=18~\mu$m, $w_{y}=8~\mu$m, $g=6~\mu$m. VO$_2$-gold hybrid structures have thickness of 200~nm. (d) Smith curves of the reflection coefficients $r$ for different VO$_2$ conductivities. Red, black, and blue curves represent underdamped, critical damped, and overdamped behaviors, respectively. The arrows indicate $f$ increasing from 0 to $\infty$.}
\label{fig:schem}
\end{figure}

Following our design philosophy, we propose a VO$_2$-gold hybrid metasurface for achieving switchable HWP/QWP, as illustrated by Figure~\ref{fig:schem}. The proposed metasurface is composed of two gold crosses bridged by a VO$_2$ short stripe, all of which stand on a gold mirror sandwiched by a polyimide spacer of thickness $h$. The metasurface can be fabricated using standard bottom-up processes. A thick gold film is first evaporated onto a substrate, followed by spin-coating of a polyimide spacer. A thin film of VO$_2$ is then deposited and is patterned using photolithography and etching. Finally, the gold crosses are patterned through photolithography, evaporation and lift-off processes.

The polarization-dependent reflection resonance characteristics as well as the near-field distributions were numerically simulated using the frequency-domain solver in CST Microwave Studio. The reflection coefficients were obtained with $r_{xx}=S_{11}$ and $r_{yy}=S_{22}$. We modelled gold using the lossy metal model with conductivity of $4.56\times 10^7$~S/m, and set the relative permittivity of the polyimide to be $\varepsilon_{\rm PI}=3.5\times(1+0.0027i)$. We adopted Drude model to describe the frequency-dependent permittivities of VO$_{2}$ \cite{wang2017hybrid}: $ \varepsilon(\omega)=\varepsilon_{\infty}- \omega_{\rm p}^{2}(\sigma)/(\omega^{2}+i\gamma\omega)$, where $\varepsilon_{\infty}$ is the permittivity at high frequency limit, $\omega_{\rm p}(\sigma) = \sqrt{\sigma/\sigma_0} \omega_{\rm p}(\sigma_{0})$ is the conductivity-dependent plasmon frequency, $\sigma$ is the conductivity, $\omega_{\rm p}(\sigma_0)= 1.4\times10^{15}$~rad/s for $\sigma_{0}=3\times10^5$~S/m, and $\gamma=5.75\times 10^{13}$~rad/s is the collision frequency. 

\begin{figure}[htbp]
\centering
\includegraphics[width=8.8cm]{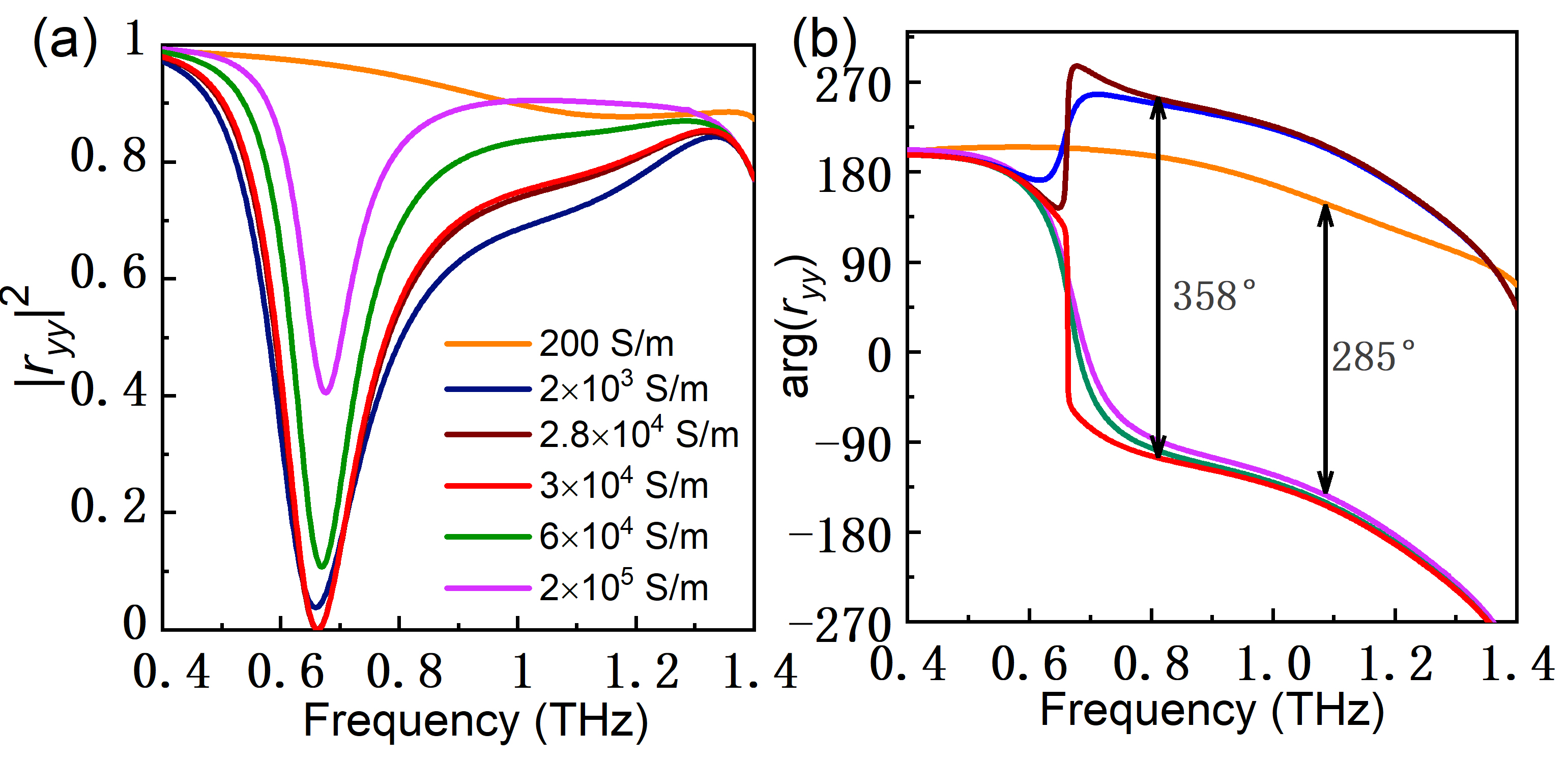}
\caption{(a) Reflectance $|r_{yy}|^2$ and (b) reflection phase ${\rm arg} (r_{yy})$ spectra for different VO$_2$ conductivities.}
\label{fig:vssigma}
\end{figure}
We first examine the dynamic transition from the overdamped to the underdamped regime for $r_{yy}$, as the VO$_2$ conductivity increases from $\sigma=200$~S/m to $2\times10^5$~S/m. Fig.~\ref{fig:vssigma}(a) shows that for $\sigma=200$~S/m, the reflectance slightly decreases as frequency increases. As the conductivity increases, a pronounced reflectance dip around 0.67~THz first becomes deeper and closer to zero reflectance, and then becomes shallower. Correspondingly, the reflectance phase variation is less than 180$^\circ$ across the resonance frequency for small conductivities of $\sigma \lesssim 2.8\times 10^4$~S/m. In other words, $r_{yy}$ is in the overdamped regime with $\Gamma_{\rm rad} < \Gamma_{\rm abs}$ and the Smith curve not enclosing the origin \cite{PRL2015ZhouLei,PRX2015ZhouLei}, as illustrated by Fig.~\ref{fig:schem}(d). For larger conductivities, the phase variation undergoes a full 360$^\circ$ range. This is because the resonance is now in the underdamped regime, for which $\Gamma_{\rm rad} > \Gamma_{\rm abs}$ and the Smith curve encloses the origin \cite{PRL2015ZhouLei,PRX2015ZhouLei}. A transition from the overdamped to the overdamped regime occurs at $\sigma \approx 2.8\times 10^4$~S/m. In this scenario, the reflectance at the resonance frequency is close to zero and the phase variation changes abruptly from less than 180$^\circ$ to 360$^\circ$.

Remarkably, the dynamic reflectance phase modulation can reach as high as 358$^\circ$ if the VO$_2$ conductivity varies from $\sigma=2\times 10^3$~S/m to $3\times 10^4$~S/m, only by 15 times. If we vary $\sigma$ by three orders of magnitude, from 200~S/m to $2\times 10^5$~S/m, the phase modulation reaches 285$^\circ$, which is very close to 270$^{\circ}$ as required for switching an HWP to a QWP according to Eq.~(\ref{eq:ryy}). More interestingly, in Fig.~\ref{fig:vssigma} we notice that, within the wide spectral band of 0.8--1.2~THz, $|r_{yy}|^2 \approx 0.9$ for both $\sigma=200$~S/m and $\sigma=2\times 10^5$~S/m, and their phase variations across the resonance frequency are nearly parallel to each other, corresponding to nearly constant phase modulation of $285^\circ$. This suggests broad bandwidth for the designed switchable HWP/QWP.

\begin{figure}[htbp]
\centering
\includegraphics[width=8.8cm]{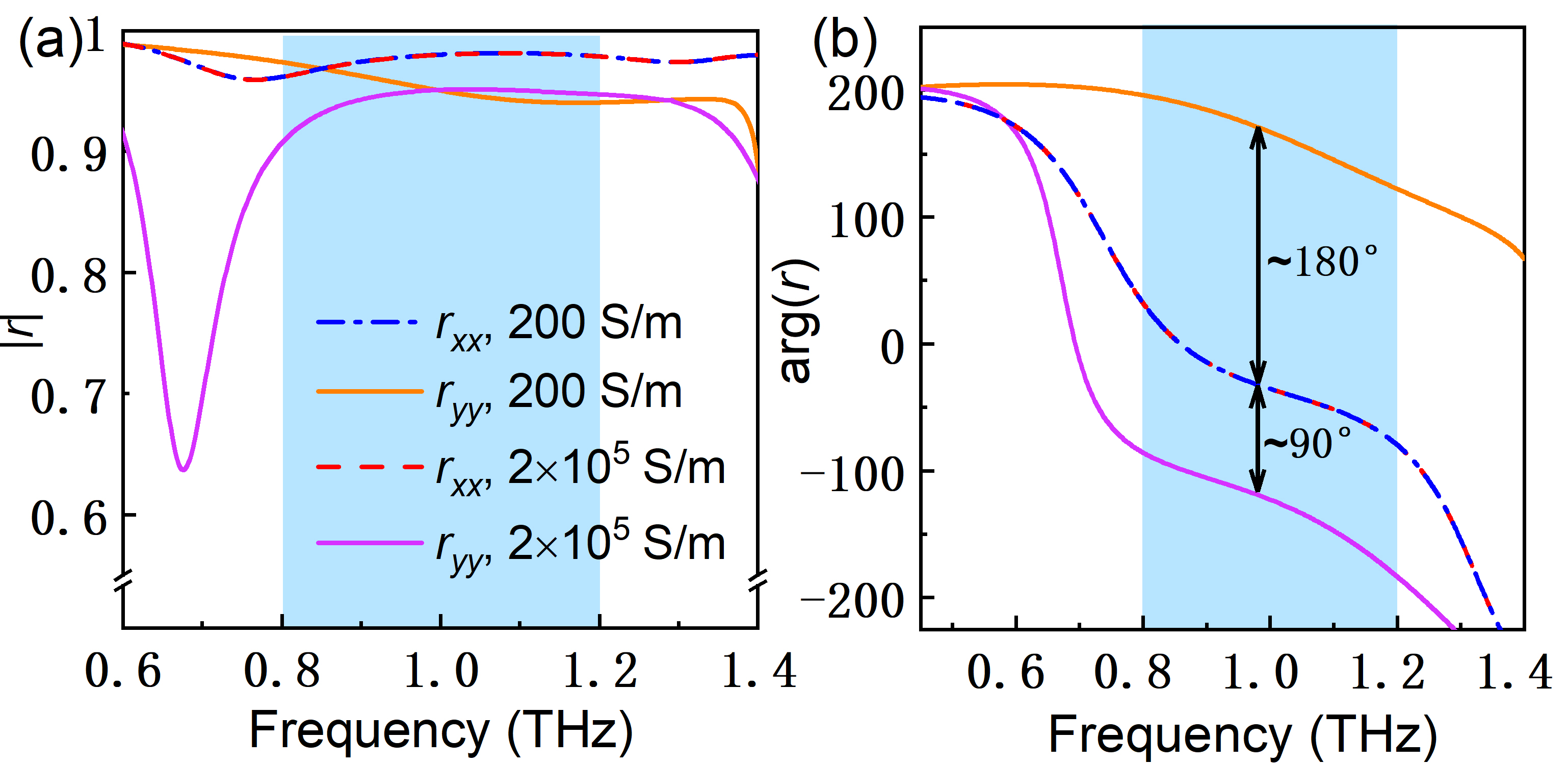}
\caption{Reflection (a) amplitude and (b) phase spectra for $r_{xx}$ and $r_{yy}$ with $\sigma=200$~S/m and $2\times 10^5$~S/m. Blue region suggests the bandwidth of the switchable HWP/QWP.}
\label{fig:switch}
\end{figure}

Having verified Eq.~(\ref{eq:ryy}), we now examine whether Eqs.~(\ref{eq:HWP}) and (\ref{eq:rxx}) can also be satisfied. Fig.~\ref{fig:switch} shows that for both $\sigma=200$~S/m and $\sigma=2\times 10^5$~S/m, both the amplitude and the phase of $r_{xx}$ remain unchanged, and $|r_{xx}|\approx 0.98$, consistent with our design philosophy. There exists a weak reflectance dip around 0.78~THz for $r_{xx}$, and the corresponding phase variation across the resonance frequency covers the full range of 360$^\circ$. This indicates that the resonance of $r_{xx}$ is in the underdamped regime. By carefully optimizing the geometric parameters of the proposed metasurface, we can tune the phase variation across the resonance frequency for $r_{xx}$ such that it locates between those for $r_{yy}$ with $\sigma=200$~S/m and with $\sigma=2\times 10^5$~S/m, and that the phase differences within 0.8--1.2~THz are $\sim 180^\circ$ for $\sigma=200$~S/m, and $\sim 90^\circ$ for $\sigma=2\times 10^5$~S/m, as shown by Fig.~\ref{fig:switch}(b). In other words, Eqs.~(\ref{eq:HWP}) and (\ref{eq:QWP}) are approximately satisfied over a broad band. Therefore, we expect that, by varying the VO$_2$ conductivity between $\sigma=200$~S/m and $2\times 10^5$~S/m, the function of the designed metasurface can be dynamically switched between an HWP and a QWP within the broad band of 0.8--1.2~THz.

\begin{figure}[htbp]
\centering
\includegraphics[width=\linewidth]{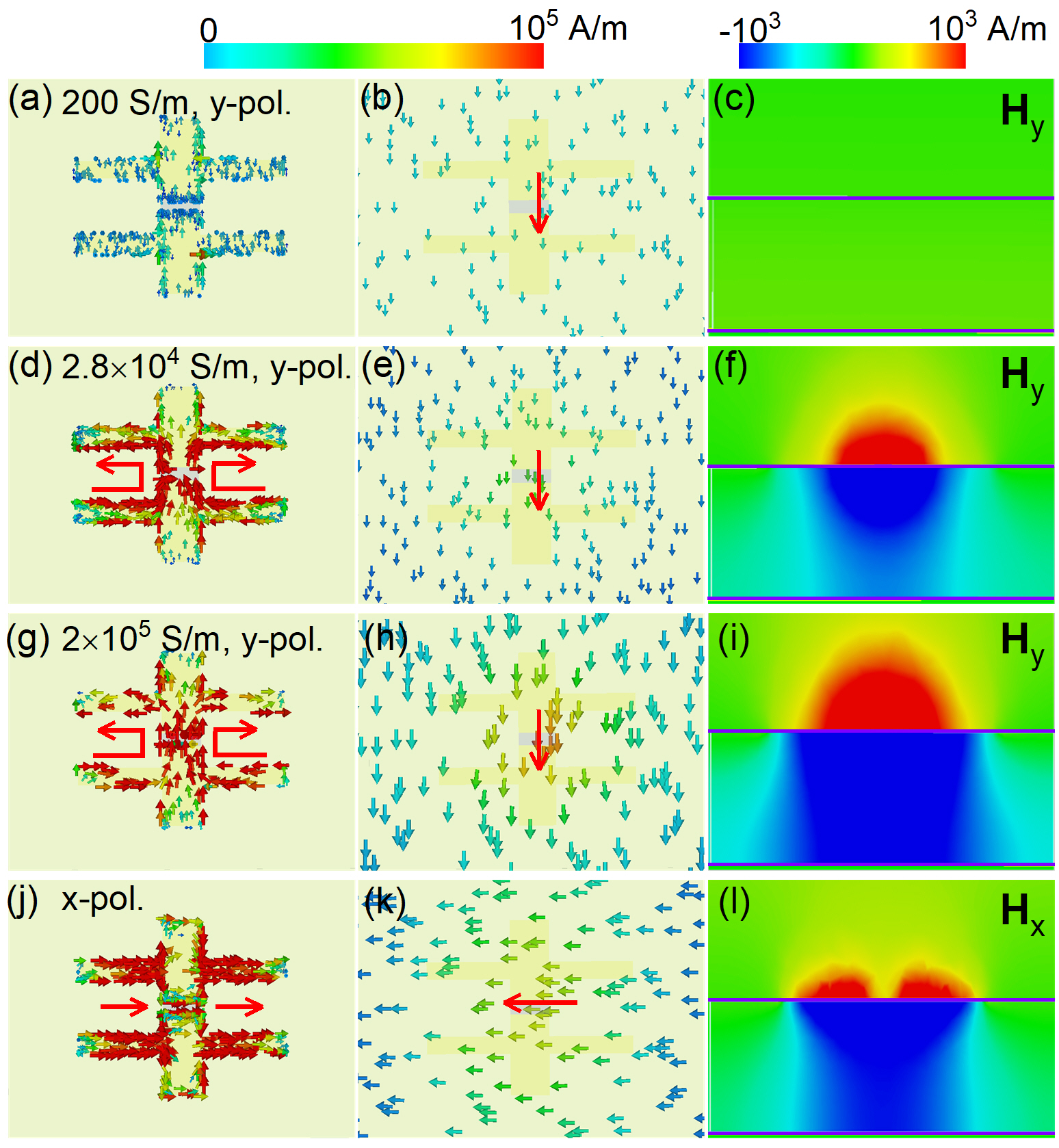}
\caption{From left to right: surface current maps on the hybrid structures and the gold film, and magnetic field maps $H_x$ in the $y-z$ plane with the spacer layer outlined by purple lines. From top to bottom: $\sigma=200$~S/m, $2.8\times10^4$~S/m, and $2\times10^5$~S/m under $\vec{E} \parallel \hat{y}$ polarization at 0.67~THz, and $\sigma=200$~S/m or $2\times 10^5$~S/m under $\vec{E} \parallel \hat{x}$ polarization at 0.78~THz.}
\label{fig:EHfields}
\end{figure}

To understand the distinct resonance characteristics of the proposed metasurface with different VO$_2$ conductivities and under the two orthogonal polarizations, we plot the surface current maps and the magnetic field distributions in Fig.~\ref{fig:EHfields}. Under the $\vec{E} \parallel \hat{y}$ polarization at $f=0.67$~THz, the surface currents and the magnetic fields are very weak for $\sigma=200$~S/m, consistent with the indistinct resonance in the reflectance spectra in Fig.~\ref{fig:vssigma}(a). For $\sigma=2.8\times 10^4$~S/m, there exist a clockwise and an counterclockwise currents on the VO$_2$-gold hybrid structures' top surface. The magnetic field is mainly confined to the hybrid structures. This corresponds to the critical transition point between the overdamped and underdamped regimes. For $\sigma=2\times10^5$~S/m, the clockwise and counterclockwise currents are relatively weak, whereas the upward-propagating surface currents on the hybrid structures and those propagating downwards on the gold film are strong. This corresponds to extremely strong magnetic field in the overcoupled MIM cavity. Therefore, as $\sigma$ increases from 200~S/m to $2\times10^5$~S/m, $r_{yy}$ undergoes a transition from the overdamped to the underdamped regime. However, under the $\vec{E} \parallel \hat{x}$ polarization at $f=0.78$~THz, the surface currents mainly propagate rightwards on the hybrid structures and leftwards on the gold film, and the corresponding magnetic field is strong in the overcoupled MIM cavity, irrespective of whether $\sigma=200$~S/m or $2\times10^5$~S/m. In other words, $r_{xx}$ stays in the underdamped regime for both conductivities.

We now rotate the sample along the $z$-axis by an angle of 45$^\circ$, as illustrated by Fig.~\ref{fig:schem}(c), and simulate the polarization conversion performance of the obtained switchable HWP/QWP. We adopt the polarization conversion ratio, ${\rm  PCR} \equiv |r_{vu}|^2/(|r_{vu}|^2+|r_{uu}|^2)$, and the degree of linear polarization, ${\rm DoLP} = \sqrt{\left(|r_{uu}|^2-|r_{vu}|^2)^2+(2|r_{vu}||r_{uu}|\cos(\Delta\Phi)\right)^2} / (|r_{vu}|^2+|r_{uu}|^2)$ to quantify the HWP, and adopt the ellipticity $\chi = S_{3}/S_{0}$ to quantify the QWP. Here, $\Delta \Phi = {\rm arg}(r_{vu}) - {\rm arg}(r_{uu})$ is the phase difference, $S_3 = 2|r_{vu}||r_{uu}|\sin(\Delta\Phi)$ and $S_0 = |r_{vu}|^2+|r_{uu}|^2$ are the Stokes parameters.

\begin{figure}[htbp]
\centering
\includegraphics[width=8.8cm]{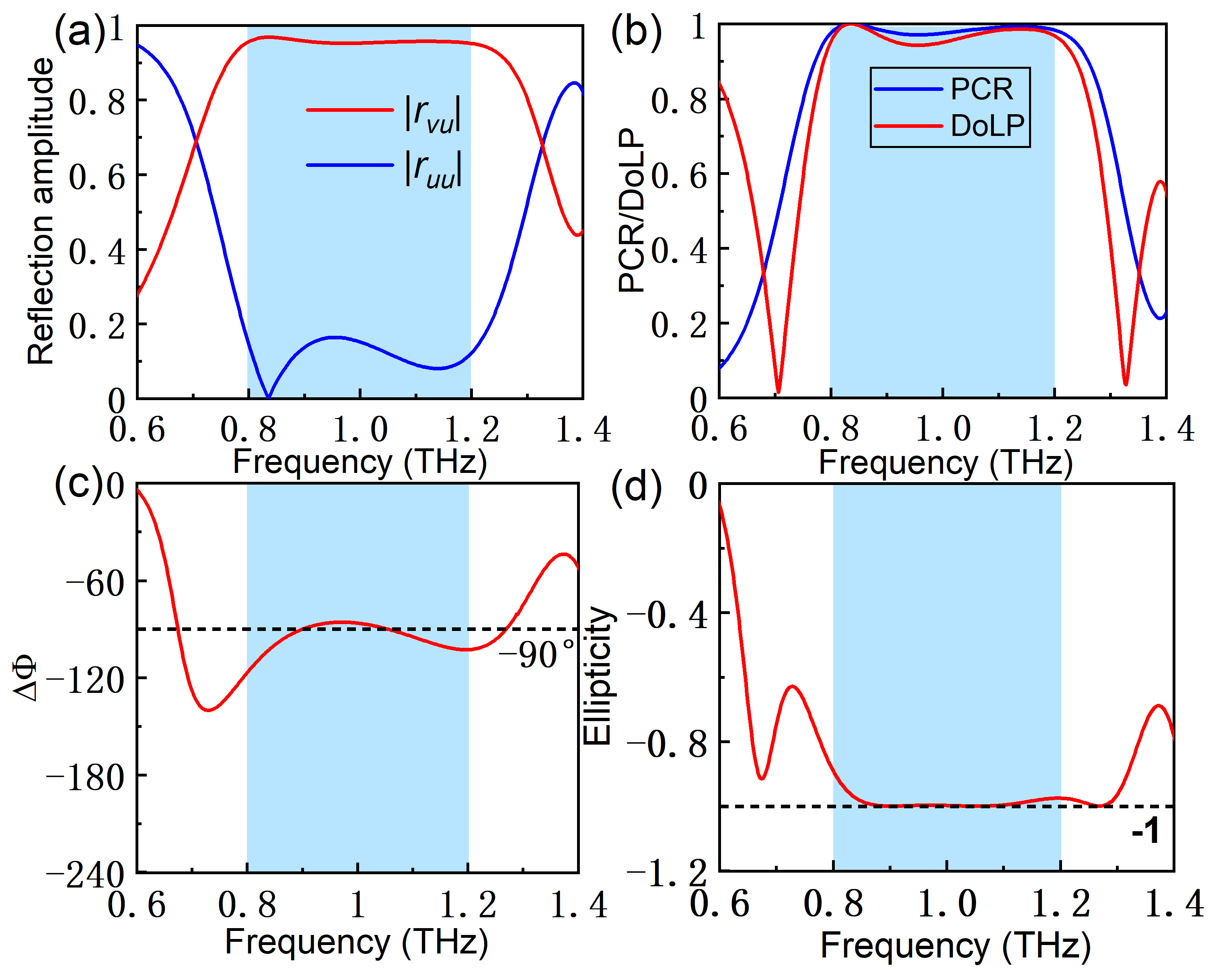}
\caption{(a) Simulated spectra of co- and
cross-polarization reflectance amplitudes, and (b) of PCR and DoLP for the HWP obtained when $\sigma= 200$~S/m. (c) Simulated spectra of phase differences and (c) of ellipticity for the QWP obtained when $\sigma= 2\times 10^5$~S/m. Blue regions define the working bandwidth.}
\label{fig:perform}
\end{figure}

When the proposed metasurface acts as an HWP for $\sigma=200$~S/m, Fig.~\ref{fig:perform}(a) shows that the cross-polarization reflection amplitude $|r_{vu}|\approx 0.96$, whereas the co-polarization response $|r_{uu}|<0.2$ within 0.8--1.2~THz. Correspondingly, ${\rm PCR}>0.96$ and ${\rm DoLP}>0.94$, as shown by Fig.~\ref{fig:perform}(b). These imply that the incident $u$-polarization wave is efficiently converted to the reflective $v$-polarization wave. When the proposed metasurface acts as a QWP for $\sigma=2\times 10^5$~S/m, Figs.~\ref{fig:perform}(c)(d) show that, the phase difference is around 90$^\circ$ and the ellipticity is close to $-1$ within the same spectral band. This suggests that the incident $u$-polarization wave is efficiently converted to the reflective left-handed circularly polarized wave. Both the HWP and the QWP share the same band with the central frequency locating at 1~THz and the relative bandwidth reaching 40\%, which is four times of the metasurface in \cite{OL2019HWP2QWP_GSi}. Therefore, the proposed metasurface can function as a broadband and high-efficiency HWP or QWP by varying the VO$_2$ conductivity.

In conclusion, we have proposed a novel design philosophy and a VO$_2$-metal metasurface for achieving a broadband switchable terahertz HWP/QWP. Results have shown that by varying the VO$_2$ conductivity from 200~S/m to $2\times10^5$~S/m, $r_{xx}$ remains in the underdamped regime whereas $r_{yy}$ undergoes a transition from the overdamped to the underdamped regime. Therefore, when $\sigma =200$~S/m, $r_{xx}$ can exhibit a 360$^\circ$ reflection phase coverage as frequency varies and thus Eq.~(\ref{eq:HWP}) can be satisfied, resulting in a high-performance HWP; when $\sigma$ raises to $2\times 10^5$~S/m, $r_{xx}$ is unchanged but $r_{yy}$ obtains phase modulation of $\sim 270^\circ$, satisfying Eqs.~(\ref{eq:rxx}) and (\ref{eq:ryy}) and leading to a high-performance QWP. Simulations results have shown that for the obtained HWP, the PCR is greater than 0.96 and the DoLP is larger than 0.94, while for the obtained QWP, the phase difference is around $-90^\circ$ and the ellipticity is close to $-1$, and that the working bandwidth of the switchable HWP/QWP covers 0.8--1.2~THz. We expect the proposed design philosophy will advance the design of reconfigurable metasurfaces for switchable functionalities beyond the terahertz regime.

\section*{Acknowledgments}
This work was supported by Shenzhen Fundamental Research and Discipline Layout project (JCYJ20180507182444250) and China Postdoctoral Science Foundation (2020M682984).

\bibliographystyle{unsrt}
\bibliography{sample}

\begin{thebibliography}{10}

\bibitem{FPC2010PolMM_rev}
Jia-ming Hao, Min Qiu, and Lei Zhou.
\newblock Manipulate light polarizations with metamaterials: From microwave to
  visible.
\newblock {\em Front. Phys. China}, 5(3):291--307, 2010.

\bibitem{RPP2016Metasurface_Rev}
Hou-Tong Chen, Antoinette~J Taylor, and Nanfang Yu.
\newblock A review of metasurfaces: physics and applications.
\newblock {\em Rep. Prog. Phys.}, 79(7):076401, 2016.

\bibitem{AM2019ManipulMS_rev}
Shuqi Chen, Zhancheng Li, Wenwei Liu, Hua Cheng, and Jianguo Tian.
\newblock From single-dimensional to multidimensional manipulation of optical
  waves with metasurfaces.
\newblock {\em Adv. Mater.}, 31:1802458, 2019.

\bibitem{NanoP2020PolMS_rev}
Yuttana Intaravanne and Xianzhong Chen.
\newblock Recent advances in optical metasurfaces for polarization detection
  and engineered polarization profiles.
\newblock {\em Nanophotonics}, 9(5):1003--1014, 2020.

\bibitem{OE2009THzPolConvertNarrow}
Xomalin~G. Peralta, Evgenya~I. Smirnova, Abul~K. Azad, Hou-Tong Chen,
  Antoinette~J. Taylor, Igal Brener, and John~F. O’Hara.
\newblock Metamaterials for thz polarimetric devices.
\newblock {\em Opt. Express}, 17:773--783, 2009.

\bibitem{Science2013THzPolConvertBroad}
Nathaniel~K Grady, Jane~E Heyes, Dibakar~Roy Chowdhury, Yong Zeng, Matthew~T
  Reiten, Abul~K Azad, Antoinette~J Taylor, Diego~AR Dalvit, and Hou-Tong Chen.
\newblock Terahertz metamaterials for linear polarization conversion and
  anomalous refraction.
\newblock {\em Science}, 340(6138):1304--1307, 2013.

\bibitem{JPD2020TuneMM_rev}
Shuyuan Xiao, Tao Wang, Tingting Liu, Chaobiao Zhou, Xiaoyun Jiang, and Jianfa
  Zhang.
\newblock Active metamaterials and metadevices: a review.
\newblock {\em J. Phys. D: Appl. Phys.}, 53:503002, 2020.

\bibitem{Res2020TuneM_rev}
Qiong He, Shulin Sun, and Lei Zhou.
\newblock Tunable/reconfigurable metasurfaces: physics and applications.
\newblock {\em Research}, 2019:1849272, 2019.

\bibitem{MOTL2020TuneMM_rev}
Lei Bao and Tie~Jun Cui.
\newblock Tunable, reconfigurable, and programmable metamaterials.
\newblock {\em Microw. Opt. Technol. Lett.}, 62(1):9--32, 2020.

\bibitem{APL2014HWP-THZ}
Y.~Cheng, W.~Withayachumnankul, A.~Upadhyay, Yan Nie, Rong~Zhou Gong, Madhu
  Bhaskaran, Sharath Sriram, and Derek Abbott.
\newblock Ultrabroadband reflective polarization convertor for terahertz waves.
\newblock {\em Appl. Phys. Lett.}, 105:181111, 2014.

\bibitem{EPL2017HWPQWPBroad}
Shaojie Ma, Xinke Wang, Weijie Luo, Shulin Sun, Yan Zhang, Qiong He, and Lei
  Zhou.
\newblock Ultra-wide band reflective metamaterial wave plates for terahertz
  waves.
\newblock {\em EPL}, 117(3):37007, 2017.

\bibitem{OME2017HWPBroad}
Rui Xia, Xufeng Jing, Xincui Gui, Ying Tian, and Zhi Hong.
\newblock Broadband terahertz half-wave plate based on anisotropic polarization
  conversion metamaterials.
\newblock {\em Opt. Mater. Express}, 7(3):977--988, 2017.

\bibitem{LPR2014QWP-THZ}
Longqing Cong, Ningning Xu, Jianqiang Gu, Ranjan Singh, Jiaguang Han, and Weili
  Zhang.
\newblock Highly flexible broadband terahertz metamaterial quarter-wave plate.
\newblock {\em Laser Photonics Rev.}, 8(4):626--632, 2014.

\bibitem{SCIREP2016qwp-thz}
Muhammad~Tayyab Nouman, Ji~Hyun Hwang, and Jae-Hyung Jang.
\newblock Ultrathin terahertz quarter-wave plate based on split ring resonator
  and wire grating hybrid metasurface.
\newblock {\em Sci. Rep.}, 6(1):39062, 2016.

\bibitem{OL2016QWPTHZ}
TIANJING Guo and CHRISTOS Argyropoulos.
\newblock Broadband polarizers based on graphene metasurfaces.
\newblock {\em Opt. Lett.}, 41:5592--5595, 2016.

\bibitem{OE2018QWPHWP-THZ}
Wendy~SL Lee, Rajour~T Ako, Mei~Xian Low, Madhu Bhaskaran, Sharath Sriram,
  Christophe Fumeaux, and Withawat Withayachumnankul.
\newblock Dielectric-resonator metasurfaces for broadband terahertz quarter-and
  half-wave mirrors.
\newblock {\em Opt. Express}, 26(11):14392--14406, 2018.

\bibitem{OSAC2018QWP2HWP}
Wei Zhang, Jianli Jiang, Jing Yuan, Shuang Liang, Jisong Qian, Jing Shu, and
  Liyong Jiang.
\newblock Functionality-switchable terahertz polarization converter based on a
  graphene-integrated planar metamaterial.
\newblock {\em OSA Continuum}, 1(1):124--135, 2018.

\bibitem{OE2018SwitchQWP}
Yun-Yun Ji, Fei Fan, Xiang-Hui Wang, and Sheng-Jiang Chang.
\newblock Broadband controllable terahertz quarter-wave plate based on graphene
  gratings with liquid crystals.
\newblock {\em Opt. Express}, 26(10):12852--12862, 2018.

\bibitem{PTL2019TunableQWP}
Mohammad~Reza Tavakol, Babak Rahmani, and Amin Khavasi.
\newblock Terahertz quarter wave-plate metasurface polarizer based on arrays of
  graphene ribbons.
\newblock {\em IEEE Photon. Technol. Lett.}, 31:931--934, 2019.

\bibitem{OL2019HWP2QWP_GSi}
Shengnan Guan, Jierong Cheng, Tiehong Chen, and Shengjiang Chang.
\newblock Bi-functional polarization conversion in hybrid graphene-dielectric
  metasurfaces.
\newblock {\em Opt. Lett.}, 44:5683--5686, 2019.

\bibitem{SCIREP2015QWPVO2}
Dacheng Wang, Lingchao Zhang, Yinghong Gu, MQ~Mehmood, Yandong Gong, Amar
  Srivastava, Linke Jian, T~Venkatesan, Cheng-Wei Qiu, and Minghui Hong.
\newblock Switchable ultrathin quarter-wave plate in terahertz using active
  phase-change metasurface.
\newblock {\em Sci. Rep.}, 5(1):15020, 2015.

\bibitem{IEEEPJ2016QWPVO2}
Dacheng Wang, Lingchao Zhang, Yandong Gong, Linke Jian, T.~Venkatesan,
  Cheng-Wei Qiu, and Minghui Hong.
\newblock Multiband switchable terahertz quarter-wave plates via phase-change
  metasurfaces.
\newblock {\em IEEE Photon. J.}, 8:5500308, 2016.

\bibitem{OE2020HWP2QWP}
Juan Luo, Xingzhe Shi, Xiaoqing Luo, Fangrong Hu, and Guangyuan Li.
\newblock Broadband switchable terahertz half-/quarter-wave plate based on
  metal-vo 2 metamaterials.
\newblock {\em Opt. Express}, 28(21):30861--30870, 2020.

\bibitem{CPL2020QWP2HWP}
Zhao Jian-Xing, Song Jian-Lin, Zhou Yao, Liu Yi-Chao, and Zhou Jian-Hong.
\newblock Switching between the functions of half-wave plate and quarter-wave
  plate simply by using a vanadium dioxide film in a terahertz metamaterial.
\newblock {\em Chin. Phys. Lett.}, 37(6):64204, 2020.

\bibitem{AOM2018HWP2ABS}
Fei Ding, Shuomin Zhong, and Sergey~I Bozhevolnyi.
\newblock Vanadium dioxide integrated metasurfaces with switchable
  functionalities at terahertz frequencies.
\newblock {\em Adv. Opt. Mater.}, 6(9):1701204, 2018.

\bibitem{OE2020switchableAT}
Xiaoxiang Dong, Xiaoqing Luo, Yixuan Zhou, Yuanfu Lu, Fangrong Hu, Xinlong Xu,
  and Guangyuan Li.
\newblock Switchable broadband and wide-angular terahertz asymmetric
  transmission based on a hybrid metal-vo2 metasurface.
\newblock {\em Opt. Express}, 28(21):30675--30685, 2020.

\bibitem{NPGAM2018_VO2Rev}
Z.~Shao, X.~Cao, H.~Luo, and P.~Jin.
\newblock Recent progress in the phase-transition mechanism and modulation of
  vanadium dioxide materials.
\newblock {\em NPG Asia Mater.}, 10:581--605, 2018.

\bibitem{wang2017hybrid}
Shengxiang Wang, Lei Kang, and Douglas~H Werner.
\newblock Hybrid resonators and highly tunable terahertz metamaterials enabled
  by vanadium dioxide {VO$_2$}.
\newblock {\em Sci. Rep.}, 7:4326, 2017.

\bibitem{PRL2015ZhouLei}
Che Qu, Shaojie Ma, Jiaming Hao, Meng Qiu, Xin Li, Shiyi Xiao, Ziqi Miao, Ning
  Dai, Qiong He, Shulin Sun, and Lei Zhou.
\newblock Tailor the functionalities of metasurfaces based on a complete phase
  diagram.
\newblock {\em Phys. Rev. Lett.}, 115:235503, 2015.

\bibitem{PRX2015ZhouLei}
Ziqi Miao, Qiong Wu, Xin Li, Qiong He, Kun Ding, Zhenghua An, Yuanbo Zhang, and
  Lei Zhou.
\newblock Widely tunable terahertz phase modulation with gate-controlled
  graphene metasurfaces.
\newblock {\em Phys. Rev. X}, 5:041027, 2015.

\end{thebibliography}

\end{document}